\documentclass[12pt]{article}
\usepackage{graphicx,amsmath,amssymb}
\usepackage{units}

\newlength{\dinwidth}
\newlength{\dinmargin}
\newcommand{\dif}{\mathrm{d}}
\newcommand{\diff}[1]{\frac{\mathrm{d}#1}{#1}}

\def\lapproxeq{\lower .7ex\hbox{$\;\stackrel{\textstyle<}{\sim}\;$}}
\def\gapproxeq{\lower .7ex\hbox{$\;\stackrel{\textstyle>}{\sim}\;$}}
\def\be{\begin{equation}}
\def\ee{\end{equation}}
\def\bea{\begin{eqnarray}}
\def\eea{\end{eqnarray}}

\def\sh{\hat s}
\def\sh2{{\hat s}^2}

\begin{document}


\begin{flushright}
IPPP/13/34  \\
DCPT/13/68 \\
\today \\
\end{flushright} 

\vspace*{0.5cm}

\begin{center}

{\Large \bf Unintegrated parton distributions in nuclei}

\vspace*{1cm}

E.G. de Oliveira$^{a,b}$, A.D. Martin$^a$, F.S. Navarra$^b$ and M.G. Ryskin$^{a,c}$  \\

\vspace*{0.5cm}

$^a$ Institute for Particle Physics Phenomenology, University of Durham, Durham, DH1 3LE \\
$^b$ Instituto de F\'{\i}sica, Universidade de S\~{a}o Paulo, C.P.
66318,05315-970 S\~{a}o Paulo, Brazil \\

$^c$ Petersburg Nuclear Physics Institute, NRC Kurchatov Institute, Gatchina, St.~Petersburg, 188300, Russia

\vspace*{1cm}

\begin{abstract}
We study how unintegrated parton distributions in nuclei can be calculated from the corresponding integrated partons using the EPS09 parametrization. The role of nuclear effects is presented in terms of the ratio $R^A=\text{uPDF}^A/A\cdot \text{PDF}^N$ for both large and small $x$ domains.\\
\end{abstract}

\vspace*{0.5cm}

\end{center}

\section{Introduction}

Distributions unintegrated over the parton transverse momentum, $k_t$, are known to be an effective way to describe hard processes in which the transverse momentum of final particle (prompt photon\cite{sal,kim}, $W,Z$ boson or Drell-Yan lepton pair\cite{wz}, heavy quark\cite{bb}, etc.) is measured. For the interaction with a heavy nucleus the momentum distribution of secondaries is affected both by the `final state' rescattering of the secondary particle inside the nuclear medium and by the nuclear modification of the incoming Parton Distribution Functions (PDFs).

In \cite{MRW} a prescription was proposed which allows the unintegrated PDF (uPDF) to be obtained from the conventional integrated PDF with NLO accuracy. This opens up the possibility to study the nuclear modifications of uPDFs based on the existing parton distributions in nucleus. Here the unintegrated distributions are calculated using the EPS09\cite{eps} nuclear parton integrated distribution functions for the case of $A=208$, lead.

Different approaches have been used to determine the unintegrated nuclear distributions~\cite{BM,rcBKA}. 
In the earlier work~\cite{BM}, the nuclear effects in the gluon distribution were studied in the small $x$ domain based on the analytical asymptotic solution of the Balitsky--Kovchegov equation~\cite{BK}. The BK equation is a non-linear evolution equation in the variable $x$ and includes effects from saturation~\cite{GLR}.

In Ref.\ \cite{rcBKA}, a Monte Carlo was developed to determine the nuclear unintegrated gluon distribution, in which, in impact parameter space, nucleons of finite radius are placed at random in positions  inside the nucleus. When the nucleus is probed at an impact parameter that is contained by only one nucleon, the unintegrated gluon distribution from free proton is used, derived from  the numerical solution of the running coupling BK equation~\cite{rcBK} starting from some initial saturation scale $Q_{s0}^2$. If the nucleus is probed at an impact parameter domain where $n$ nucleons overlap, then the initial saturation scale is multiplied $n$, i.e., $n Q_{s0}^2$.

In contrast with the previous studies, the present work covers the whole region of $x$ up to $x=1$ and,  besides the screening corrections, accounts for the Fermi motion, EMC effect and the antishadowing at not too low $x$. Also, all species of unintegrated partons are obtained, including unintegrated quark distributions. This work   is based in Dokshitzer--Gribov--Lipatov--Altarelli--Parisi evolution~\cite{DGLAP} in which $\ln  Q^2$ terms are properly taken into account at each order, and the unintegrated distributions are functions of three variables, $x$, $k_t$, and $Q^2$, instead of just the first two. DGLAP provides a linear evolution and saturation effects are absent, except in the parametrization of the distributions at the starting scale. In addition, our approach includes the suppression of unintegrated  distributions at small $k_t$ caused by the Sudakov factor.

\section{Nuclear modification}

Recall that a parton distribution in a nucleus (PDF$^A$) is not equal to the sum of the PDFs in the component nucleons ($A\cdot$PDF$^N$)~\footnote{Strictly speaking as a `reference quantity' in denominator we have to use not $A\cdot$PDF$^N$ but $Z\cdot\mbox{PDF}^p+N\cdot\mbox{PDF}^n$. However this does not matter since below we consider the flavour singlet parton distributions only.}. There are different physical effects depending on the value of $x$. We summarize here in terms of ratio
\be
R^A \equiv \frac{\text{PDF}^A}{A\cdot \text{PDF}^N}
\ee
as follows:

\begin{itemize}
\item At very small $x$ the parton density is smaller (than the simple sum) due to absorptive (shadowing) effects. 

\item At larger $x$, $0.03 \lesssim x \lesssim 0.1$,  the  value of PDF$^A$ exceeds the sum, $A\cdot$PDF$^N$. This antishadowing is just due to momentum conservation. After the fusion of two parton cascades (originating from two different nucleons) into the one branch of the parton cascade we get a lower number of low $x$ partons but the momentum fraction, $x$, carried by each parton becomes larger just near the fusion position.

\item Next, for $x \gtrsim 0.3$ there is an EMC effect\cite{EMC} and the ratio $R^A$ becomes less than 1.

\item Finally, at very large $x \gtrsim 0.8$, we have an enhancement $R^A>1$ due to Fermi motion.
\end{itemize}

The unintegrated parton distributions are obtained from \cite{MRW}:
\begin{equation} \label{eq:uPDF}
  f_a(x,k_t^2,\mu^2)~=~\int^1_x\!\mathrm{d}z\;T_a(k^2,\mu^2)\;
  \frac{\alpha_S(k^2)}{2\pi}\sum_{b=q,g} \tilde{P}_{ab}\left(z,\Delta\right)\, \frac{x}{z} b \left(\frac{x}{z},k^2\right)\Theta (1-z-k^2_t/\mu^2),
\end{equation}
where $k^2=k^2_t/(1-z)$ and $b \left(x,k^2\right)$ is the integrated parton distribution, for example, $b(x, \mu^2) = g (x, \mu^2)$. The cutoff $\Delta$ in $z$ integration is specified below, see (\ref{eq:delta1}). The Sudakov factor $T_a (k^2, \mu^2)$ resums the virtual DGLAP contributions during the evolution from $k^2$ to $\mu^2$. It is given by:
\begin{equation}
  T_a(k^2,\mu^2) = \exp\left(-\int_{k^2}^{\mu^2}\!\diff{\kappa^2}\,\frac{\alpha_S(\kappa^2)}{2\pi}\,
    \int_0^1\!\dif{\zeta }\;~\zeta \sum_b \tilde{P}_{ba}(\zeta,\Delta) \right).
  \label{eq:Sud}
\end{equation}

The tilde splitting functions are given by the usual $\alpha_s$ expansion $\tilde{P}=\tilde{P}^{(0)}+(\alpha_S/2\pi) \tilde{P}^{(1)} + \dots$ and they are defined from the unregulated DGLAP splitting kernels. For non-diagonal elements, one has $\tilde{P}(x,\Delta) = P(x)$, while for diagonal elements:
\begin{equation}
  \tilde{P}^{(i)}_{aa}(x,\Delta) = P_{aa}^{(i)}(x) - \Theta(z-(1-\Delta)) F^{(i)}_a p_{aa}(x)
\end{equation}
with $F_q^{(0)} = C_F$, $F_g^{(0)} = 2 C_A $, and 
\begin{align}
F_a^{(1)} &= - F_a^{(0)} \left( T_R N_F \frac{10}{9} + C_A \left( \frac{\pi^2}{6} - \frac{67}{18} \right) \right).
\end{align}
Also, 
\begin{align}
p_{qq}(x) &= \frac{1+x^2}{1-x} \\
p_{gg}(x) &= \frac{x}{1-x} + \frac{1-x}{x}  + x(1-x).
\end{align}

The cutoff in $z$, $\Delta$, accounts for the coherence of gluon radiation amplitudes which leads to the angular ordering of emitted gluons. It is function of transverse momentum and the scale $\mu$. In (\ref{eq:uPDF}) it reads
\begin{equation} \label{eq:delta1}
\Delta=\frac{k_t}{\mu+k_t},
\end{equation}
while in (\ref{eq:Sud}), being written in terms of virtuality,
\begin{equation} \label{eq:delta2}
\Delta=\frac{\sqrt{\kappa^2(1-\zeta)}}{\mu+\sqrt{\kappa^2(1-\zeta)}} 
\qquad \Longrightarrow \qquad
\Delta = \frac{2\kappa^2}{2\kappa^2 + \mu^2 + \sqrt{4\kappa^2\mu^2 + \mu^4}}.
\end{equation}
We use NLO kinematics throughout.

At first sight, the unintegrated parton distributions, $f(x, k_t,\mu)$ should have  the same behaviour as the parent integrated parton $b(x/z,k^2)$ in (\ref{eq:uPDF}). However, the integral (\ref{eq:uPDF}) samples the parton density at a larger $x\to x/z$ and at a somewhat different scale $k^2=k^2_t/(1-z)$. Therefore the uPDFs become shifted to the left, that is to a smaller values of both $x$ ($x<x/z$) and  $k_t$ ($k^2_t<k^2$). This leads to the distortion of the nuclear modification effects. Depending on the particular kinematics in some regions, say $x \sim 0.01$, we may get $R^A<1$ (shadowing) at low $k_t$ and antishadowing, $R^A>1$ at larger $k_t$.

\section{Results}

\vspace{-0.0cm}
\begin{figure} [htb]
\begin{center}
\includegraphics[height=6.8cm]{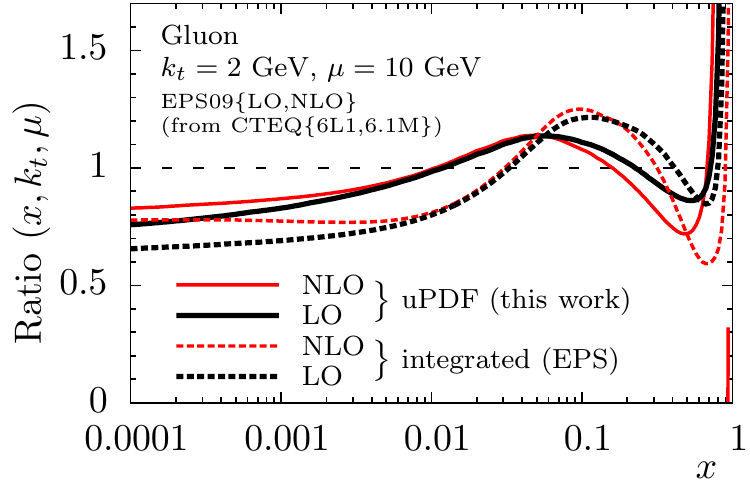}
\includegraphics[height=6.8cm]{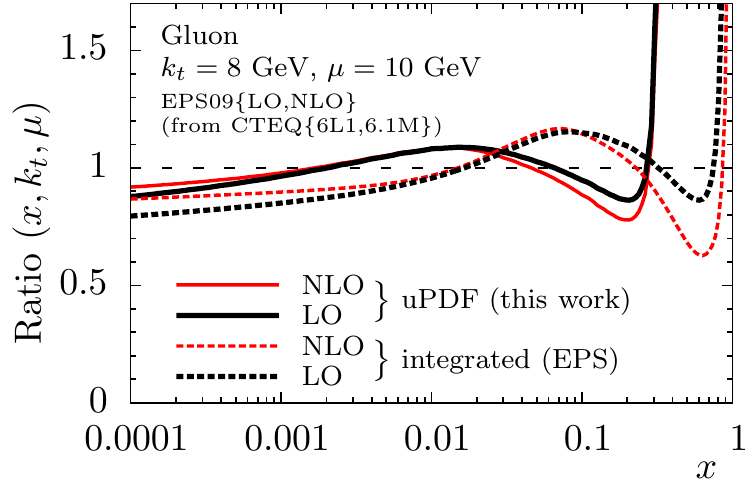}
\vspace*{-0.5cm}  
\caption{\sf Gluon ratios obtained using EPS09 nuclear PDFs with $\mu=10$ GeV. The continuous curves are the unintegrated ratios uPDF$^A / A\cdot\text{uPDF}^N$. The integrated ratios are given simply by EPS09\{LO,NLO\}$(x,k_t)$ / CTEQ\{6L1,6.1M\}$(x,k_t)$. }
\label{fig:ug}
\end{center}
\end{figure} 
\vspace{-0.0cm}
\begin{figure} [htb]
\begin{center}
\includegraphics[height=6.8cm]{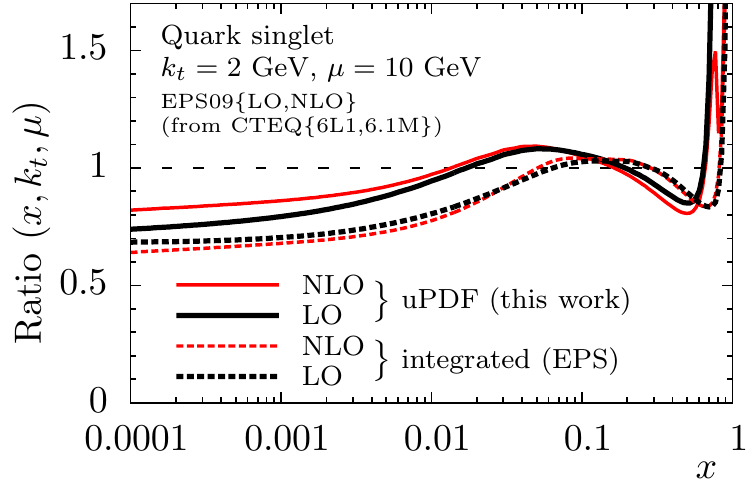}
\includegraphics[height=6.8cm]{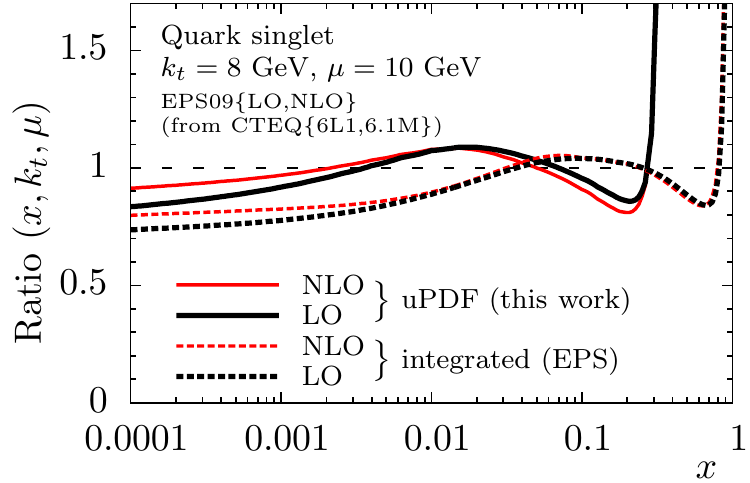}
\vspace*{-0.5cm}  
\caption{\sf Quark singlet ratios obtained using EPS09 nuclear PDFs with $\mu=10$ GeV. The continuous curves the unintegrated ratios $R^A$. The integrated ratios are given simply by EPS09\{LO,NLO\}$(x,k_t)$ / CTEQ\{6L1,6.1M\}$(x,k_t)$.}
\label{fig:uq}
\end{center}
\end{figure} 
\vspace{-0.0cm}
\begin{figure} [htb]
\begin{center}
\includegraphics[height=5.2cm]{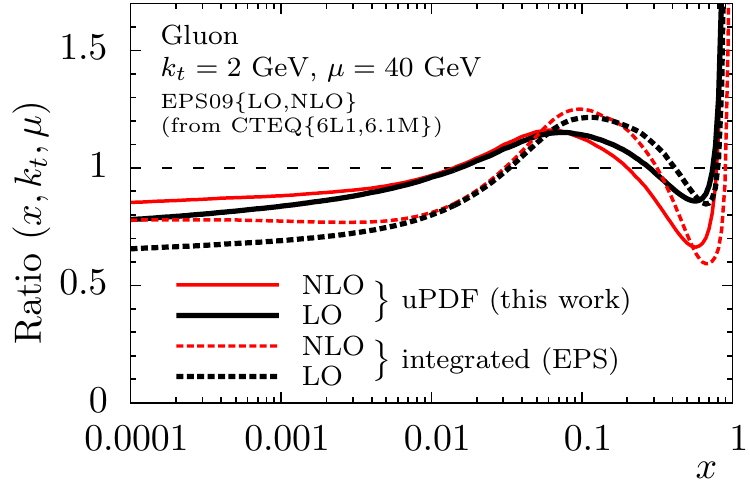}
\includegraphics[height=5.2cm]{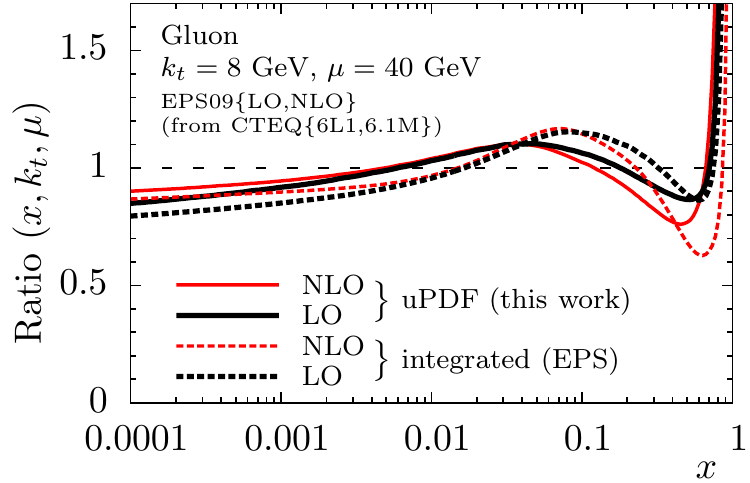}
\includegraphics[height=5.2cm]{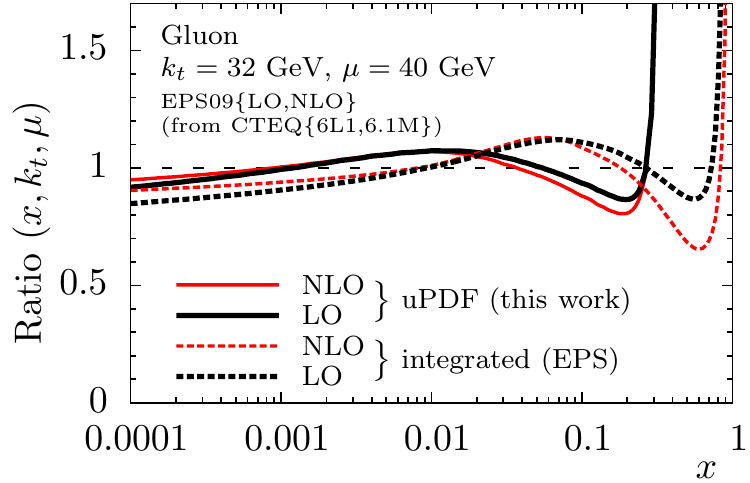}
\vspace*{-0.5cm}  
\caption{\sf Gluon ratios obtained using EPS09 nuclear PDFs  with $\mu=40$ GeV. The continuous curves are the unintegrated ratios $R^A$. The integrated ratios are given simply by EPS09\{LO,NLO\}$(x,k_t)$ / CTEQ\{6L1,6.1M\}$(x,k_t)$. }
\label{fig:ug40}
\end{center}
\end{figure} 
\vspace{-0.0cm}
\begin{figure} [htb]
\begin{center}
\includegraphics[height=5.2cm]{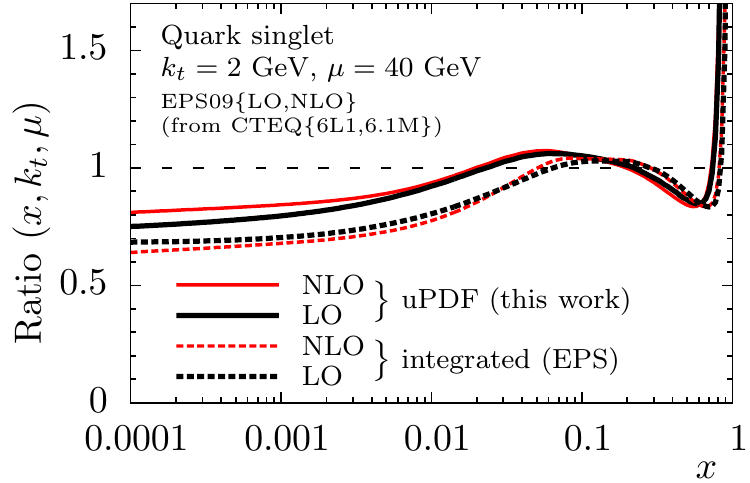}
\includegraphics[height=5.2cm]{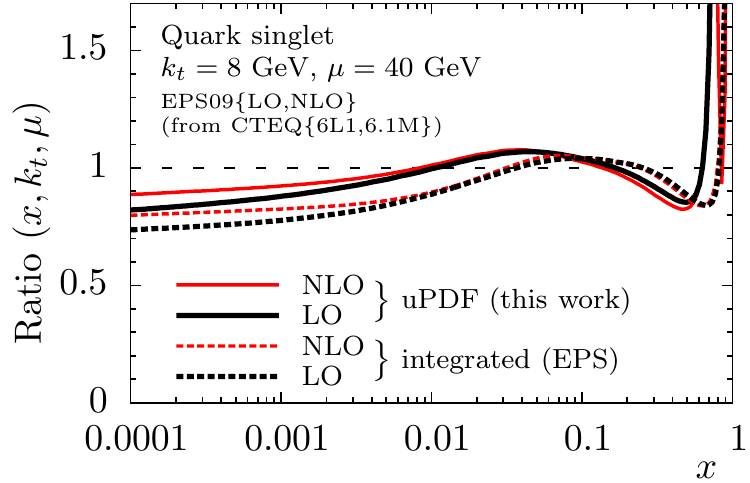}
\includegraphics[height=5.2cm]{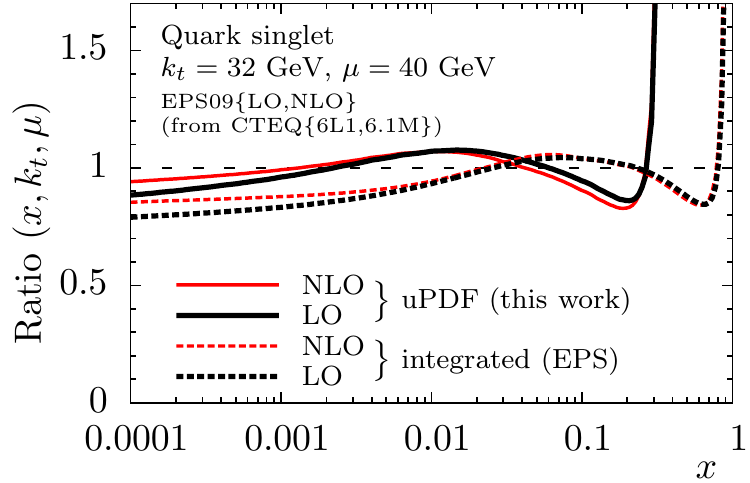}
\vspace*{-0.5cm}  
\caption{\sf Quark singlet ratios obtained using EPS09 nuclear PDFs with $\mu=40$ GeV. The continuous curves the unintegrated ratios $R^A$. The integrated ratios are given simply by EPS09\{LO,NLO\}$(x,k_t)$ / CTEQ\{6L1,6.1M\}$(x,k_t)$.}
\label{fig:uq40}
\end{center}
\end{figure}

The result of calculations are presented in Figs.\ \ref{fig:ug} and \ref{fig:uq} in the form of the ratios uPDF$^A$/$A\cdot$PDF$^N$ for the gluon and the singlet quark unintegrated distributions obtained based on the integrated EPS09 nuclear PDFs~\cite{eps}. For comparison we plot also the analogous ratio for the integrated EPS09 PDF taken at the same $x$ and the scale $k^2_t$. For the free proton baseline, the same PDFs used in the fit of EPS09 were employed here, i.e., NLO CTEQ6.1M \cite{CTEQ6.1} and LO CTEQ6L1 \cite{CTEQ6}.

We consider both the LO and the NLO prescriptions to calculate the uPDF. In both cases we account for the kinematical factor $\mu^2>k^2=k^2_t/(1-z)$ which limits the available value of $k_t<\mu\sqrt{1-z}$ in the unintegrated distributions with the fixed hard scale $\mu$. For a large $k_t$, relatively close to the value of $\mu$, these kinematics lead to a vanishing of the nucleon uPDF already at $x\sim 0.3\ - \ 0.5$. Indeed, due to  angular ordering in gluon emission we have an upper limit $z<1-\Delta=\mu/(k_t+\mu)$ in the diagonal tilde splitting functions $\tilde{P}_{aa}(z,\Delta)$ and in the integration in $z$ there is a kinematical upper limit $z<1-k_t^2/\mu^2$ enforced by the $\Theta$ function in (\ref{eq:uPDF}). On the other hand, in(\ref{eq:uPDF})  $z>x$; that is, even starting with a $\delta(1-x)$ integrated distribution, for $k_t=0.8\mu$ we get a zero uPDF for $x>0.36$. 

Actually the nucleon PDFs decrease sharply as $x\to 1$. For a heavy nucleus this distribution is washed out by Fermi motion, leading to a large ratio
\be
R^A=\frac{\text{PDF}^A}{A\cdot \text{PDF}^N}>1 
\ee
at $x$ close to 1. Since the unintegrated distribution is shifted by the kinematical inequality $z<1-\Delta$ and starts to decrease at a lower $x$, the increase of the ratio $R^A$ takes place earlier. In Figs.\ \ref{fig:ug} and \ref{fig:uq} it looks like a singularity of $R^A$, with an $x$-position which moves to smaller $x$ when $k_t$ becomes closer to $\mu$.\\

At very small $x$, absorption affects the uPDF$^A$ less than the integrated PDF$^A$. This is due to the fact that the integrated distribution, $xg(x,k^2_t)$ (or $x q(x,k^2_t$)) includes all partons with transverse momenta $k'_t<k_t$, while in uPDF we deal with partons of momentum $k_t$ only. At lower $k'_t$ the absorptive cross section, $\sigma^{abs}\propto 1/k^{'2}$, is larger, leading to stronger shadowing of the integrated distributions~\footnote{Note that at a larger $k_t$ we have a weaker shadowing in Figs.\ \ref{fig:ug} and \ref{fig:uq}}. Correspondingly in the unintegrated case we observe a weaker {\em anti}shadowing (also shifted to a smaller $x\sim 0.01\text{--}0.02$). 

The nuclear modification effects observed for the uPDF$^A$, obtained under the LO and the NLO prescriptions, are qualitatively the same. However in the NLO case, which samples some contributions beyond strong $k_t$ ordering, the difference between the shadowing of the uPDF and the integrated PDF is smaller.

In Figs.\ 3 and 4 we present the ratios $R^A$ calculated at a larger scale $\mu=40$ GeV. In the low $x$ region the nuclear modification is controlled by the value of $k_t$ and practically is independent of $\mu$; see, for example, the comparison of results in Figs.\ \ref{fig:ug} and \ref{fig:ug40} for fixed $k_t= 2 $ or 8 GeV (and again the comparison of Figs.\ \ref{fig:uq} and \ref{fig:uq40}). In (\ref{eq:uPDF}) the value of $\mu$ affects only the Sudakov factor $T(k^2,\mu^2)$ which is almost exactly canceled in the ratio $R^A$. On the other hand, for large $x$ it becomes crucial that the scale $\mu$ determines the limits of the $z$ integration and therefore the shift between the $R^A$ curves for integrated and unintegrated distribution depends on the ratio $k_t/\mu$. Indeed, we observe practically the same shift in $x$ for the cases of \{$\mu=40$ GeV, $k_t=32$ GeV\} and  \{$\mu=10$ GeV, $k_t=8$ GeV\}.

\section{Conclusion}

The unintegrated quark and gluon $x$-distributions for a lead nucleus were calculated for different values of transverse momenta $k_t$ based on the integrated PDFs for the nucleus~\cite{eps} using both the LO and the NLO prescriptions~\cite{kim,wz}. We present the ratios of the parton distributions for a lead nucleus to that for the sum of the free constituent nucleons. We discuss the role of the kinematical effects which (a) shift the unintegrated distribution to a smaller $x$ values, (b) wash out the distribution, and (c) lead to a weaker absorption in the uPDF$^A$ case.
 We show that the absorptive effects depend mainly on the value of $k_t$, while the shift of the $R^A$ curve in $x$ is controlled by the $k_t/\mu$ ratio.

\section*{Acknowledgements}

EGdO and MGR thank the IPPP at the University of Durham for hospitality. This work was supported by the grant RFBR 11-02-00120-a and by the Federal Program of the Russian State RSGSS-4801.2012.2; and by CNPq and FAPESP (Brazil).  EGdO is supported by FAPESP  under contract 2012/05469-4.

For a review of the EMC effect see for instance
M.~Arneodo, 
Phys.Rept. {\bf 240} (1994) 301,
 and references therein.
For recent measurement of the EMC effect see, for example:
  J.~Seely {\it et al.},
  Phys.\ Rev.\ Lett.\  {\bf 103}, 202301 (2009)
  [arXiv:0904.4448 [nucl-ex]].

\bibitem{CTEQ6} 
  J.~Pumplin, D.~R.~Stump, J.~Huston, H.~L.~Lai, P.~M.~Nadolsky and W.~K.~Tung,
  JHEP {\bf 0207}, 012 (2002)  [hep-ph/0201195].

\bibitem{CTEQ6.1} 
  D.~Stump, J.~Huston, J.~Pumplin, W.~-K.~Tung, H.~L.~Lai, S.~Kuhlmann and J.~F.~Owens,
  JHEP {\bf 0310}, 046 (2003)  [hep-ph/0303013].

\end{document}